\documentclass[a4]{IEEEtran}
\usepackage{cite}
\usepackage{amsmath,amssymb,amsfonts}
\usepackage{algorithmic}
\usepackage{graphicx}
\usepackage{textcomp}
\usepackage{xcolor}
\usepackage{multirow, booktabs}
\usepackage{listings,multicol}
\usepackage{xfrac}
\usepackage[utf8]{inputenc}
\usepackage[T1]{fontenc}
\usepackage{fixltx2e}
\usepackage{csquotes}
\usepackage{amsmath,diagbox,tabu}
\usepackage{comment}

\usepackage{siunitx}
\usepackage{pgfplots}

\begin{document}

\title{SklCoin: Toward a Scalable Proof-of-Stake and Collective Signature Based Consensus Protocol for Strong Consistency in Blockchain}
\author{\IEEEauthorblockN{Zakwan Jaroucheh, Baraq Ghaleb, William J Buchanan}
\IEEEauthorblockA{\\
School of Computing, Edinburgh Napier University, Edinburgh, UK\\
z.jaroucheh@napier.ac.uk, b.ghaleb@napier.ac.uk, b.buchanan@napier.ac.uk}\\
}

\maketitle
\begin{abstract}
The proof-of-work consensus protocol suffers from two main limitations: waste of energy and offering only probabilistic guarantees about the status of the blockchain. This paper introduces SklCoin, a new Byzantine consensus protocol and its corresponding software architecture. This protocol leverages two ideas: 1) the proof-of-stake concept to dynamically form stake-proportionate consensus groups that represent block miners (stakeholders), and 2) scalable collective signing to efficiently commit transactions irreversibly. SklCoin has immediate finality characteristic where all miners instantly agree on the validity of blocks. In addition, SklCoin supports high transaction rate because of its fast miner election mechanism.
\end{abstract}

\IEEEpeerreviewmaketitle
\section{Introduction}
Blockchain technology has recently attracted the attention of the world owing to its great potential in realizing a plethora of applications where trust is a decisive element. In fact, Blockchain has eliminated the need for centralized third-party verification. Instead, the centralized approach has been replaced by a decentralized form of trust governed by multiple entities (nodes) \cite{Un-blockchain}\cite{technical-survey}. At the heart of Blockchain is the consensus algorithm where those multiple entities need to reach consensus or agreement on the state of the Blockchain (deciding which nodes are more qualified to add the next block of transactions into the blockchain and verify the results). In general, the consensus algorithms of Blockchain are classified into two main classes; proof-based and voting-based consensus. In proof-based consensus, a node is required to provide the network with some sort of proof that it is the most qualified one to add the next block. In voting-based consensus, a leader node to suggest the next block is first elected which is then voted on (i.e., the suggested block) by some other nodes and finally committed into the blockchain if it has received the vast majority of votes.

Proof-of-work (PoW) used by several cryptocurrencies including Bitcoin is an example of proof-based consensus. The PoW requires a node (miner) to solve a computationally expensive cryptographic puzzle  (i.e. finding a block hash with pre-specified leading zeros) before it can add the next block of transactions into the blockchain. A transaction processed by means of PoW is argued to be irreversible and secure as long as malicious miners control less than the half of the hashing power and the age of added block is around one hour. Key issues with the PoW includes: 1) the significant amount of power consumed by the network to find the required proof, 2) the one-hour latency in confirming transaction limits Bitcoin's suitability for real-time transactions, and 3) Bitcoin’s consensus algorithm provides only probabilistic consistency guarantees. Other alternative for a consensus protocol is to provide strong consistency where all miners instantly agree on the validity of blocks, without wasting computational power resolving inconsistencies (forks) \cite{kogias2016enhancing}. 

To address the PoW limitations, several proof-based consensus algorithms have been introduced in the literature including proof-of-stake, proof-of-elapsed time, proof-of-space among many others. The ultimate goal of such proposals was to remove the PoW computationally expensive proofs while preserving its same level of security and irreversibility. For instance, in the PoS, a miner is selected randomly based on the amount of coins it has staked into the system. Hence, the more the coins an account staked into the network, the higher the possibility of being selected to produce a block and earn the corresponding transactions fees. Theoretically, it is assumed that the probability of selecting the next miner in PoS is proportional to the account’s balance. However, it has been proved in \cite{nxtcoin} that this is not entirely accurate when the probability is only based on the amount staked. In addition,  the pseudo-random nature of PoS may generate  some new attack vectors \cite{nxtcoin}.
Another issue with proof-based consensus other than PoW is the ability of a miner to mine effortlessly in secret as there is no associated cost with mining process. To remedy the limitations of proof-based consensus, researchers have investigated the potential of employing Byzantine fault tolerant (BFT) consensus algorithms in blockchain. In BFT consensus, the process of appending the next block to the blockchain is no more the responsibility of single node or a miner. Instead, the to-be-appended block should be first agreed on and approved by a rolling committee ( i.e., a group of miners) that reaches such an agreement through rounds of  communications.

Practical Byzantine Fault Tolerance (PBFT) \cite{Castro}, was the first practical approach that allowed for Byzantine fault tolerant applications with low-overhead \cite{Understanding}. This work introduces SklCoin, a cryptocurrency based on the principles of PBFT algorithm. PBFT brings strong consistency (ensuring that clients need to wait only for the next block rather than the next several blocks, e.g. in bitcoin, to verify that a transaction has been committed). By leveraging the strong consistency of PBFT, SklCoin addresses three main challenges: (1) open membership, (2) scalability in terms of the number of nodes, and (3) transaction commitment rate.

In this paper we make the following key contributions:

\begin{itemize}
\item To address the openness issue, SklCoin introduces a new software architecture where the consensus and signing groups are determined dynamically in every time-slot as will be seen later. 
\item SklCoin implements Byzantine consensus using collective signing scheme. We adopt the idea of ByzCoin \cite{kogias2016enhancing}, to build SklCoin as PBFT atop CoSi. However, this collective signing scheme could be replaced by another scheme. 
\item Finally, SklCoin creates a new mechanism that allows all miners to agree on the slot-leader in every time-slot. This allows all miners to instantly agree on the validity of the proposed block and on the validity of the leader who proposed the block, rendering the verification process scalable.
\end{itemize}

\section{Related Work}
There have been numerous recent  studies with the goal to alleviate the shortcomings of both proof-based and Byzantine-based consensus algorithms. For instance, \cite{liu2018scalable} introduce the FastBFT consensus algorithm that integrates several technologies including lightweight secret sharing, trusted execution environments (TEEs), tree topology, optimistic execution, and failure detection, hence, realizing a low latency infrastructure.
The authors of  \cite{miller2016honey} argued, however, that weakly synchronous consensus algorithms are ill-suited for environments where messages are not guaranteed to be delivered in a certain amount of time. Hence, they introduce a leaderless consensus protocol that  uses a novel atomic broadcast protocol to guarantee liveness in  asynchronous environments. Crain et al \cite{crain2018dbft} propose another leaderless consensus protocol with the aim to overcome the problem of a faulty coordinator (i.e. leader) in asynchronous environments. The rational here is that a synchronous round can be declared completed by a process upon receiving a pre-specified threshold of messages which nullifies the need to wait for potentially slow messages from the leader. 

Overcoming the scalability issue in proof-based consensus algorithms was the focus of several research efforts. For instance, Bitcoin-NG \cite{Bitcoin-NG} proposes a new consensus algorithm that combines Byzantine fault tolerance with PoW aiming to lower the latency of transaction processing in PoW-based blockchain. The key idea is to decouple the process of miner election from transaction verification by introducing two different blocks types; Keyblocks and Microblocks. Keyblocks serve the purpose of leader selection by means of PoW. The leader then will take the responsibility of creating Microblocks of transactions that need only to be signed by the leader without the need for the power-hungry PoW. However, Bitcoin-NG is still susceptible to several issues including forking, even deliberately and history rewriting. ByzCoin \cite{kogias2016enhancing} proposes to overcoming aforementioned problems by introducing the idea of collective signing to achieve consensus in blockchain. A block is considered valid if collectively signed by a group of recently-successful block miners. However, ByzCoin inherits the power-inefficient PoW algorithm. In addition, it only gives the chance for recently successful miners to be leaders; that prevents new joining nodes from being elected.  

In Schnorr signing, we can aggregate public keys of $P$ participants into a single signing key \cite{ohta1999multi}, and uses the non-interactive version of the Fiat-Shamir heuristic \cite{Fiat-Shamir}.  Using Elliptic Curve methods, to sign a message we take a random value ($k$) and a private key value ($d$) and a generator point $G$ ($G$ is a base point on an elliptic curve) and compute:

\begin{equation}
Q=dG
\end{equation}

and:

\begin{equation}
R=kG
\end{equation}

For it to be non-interactive we then calculate:
\begin{equation}
e=H(R\parallel M)
\end{equation}

and then:
\begin{equation}
s=k-ed
\end{equation}

The signature is then $(s,e)$. To verify we compute:
\begin{equation}
r_{v}=sG+eQ    
\end{equation}

and then:

\begin{equation}
e_{v}=H(r_{v}\parallel M)
\end{equation}

We then check that $r_{v}=e_{v}$.


Each participant has a private key ($a_i$) and a public key $A_i=a_i G$. We can then determine the aggregate public key with: 

\begin{equation}
A = \sum_{i \in P} A_i
\end{equation}

\section{Background}

\subsection{Byzantine Fault Tolerance}
BFT consensus algorithms focus on building fault tolerance in the face of unreliable systems provisioning mainly for fail-stop faults. In the face of such failures, these algorithms guarantee progress and consistency in the data structures that were replicated across the nodes. The number of nodes needed in such networks are $2f + 1$ to be able to tolerate $f$ fail-stop failures. Tolerating Byzantine faults, increases the complexity of the consensus protocol by adding several extra layers of messaging into the system \cite{Solida}. Practical Byzantine Fault Tolerance (PBFT) \cite{Castro}, was the first practical approach that allowed for Byzantine fault tolerant applications with low-overhead. 

PBFT uses the concept of primary and secondary replicas, where the secondary replicas automatically check the sanity and liveness of decisions taken by the primary and can collectively switch to a new primary if the primary is found to be compromised \cite{LearnBitcoin}. PBFT brings strong consistency (ensuring that clients need to wait only for the next block rather than the next several blocks, e.g. in bitcoin, to verify that a transaction has committed). 

PBFT relies upon a primary node (leader) to begin each round and proceeds if a two-thirds quorum exists. Three distinct phases happen in every round of PBFT:
\begin{enumerate}
\item Pre-prepare phase: The current primary node (leader) announces their proposal that the nodes should agree upon. Once received, every node validates the proposal and multicasts a prepare message to the group of nodes. 
\item The nodes wait until they receive $(2 f + 1)$ prepare messages and then publish a commit message. 
\item The nodes wait until they receive $(2 f + 1)$ commit messages to make sure that enough nodes have agreed on the proposal and committed themselves to it.
\end{enumerate}

Since the leader is a potential attack vector, PBFT implements a view-change protocol that ensures liveness in the face of a faulty leader. The nodes initiate a view-change (i.e. changing the leader) if they detect either malicious behavior or an unsatisfactory progress. If a quorum of $(2 f + 1)$ nodes agrees that the leader is faulty, then a next leader takes over.

PBFT has the following limitations:
\begin{itemize}
\item PBFT was not designed for scalability in large consensus group of nodes. Each PBFT node normally communicates directly with every other nodes during each consensus round, resulting in $O(n^2)$ communication complexity.
\item PBFT normally assumes a well-defined, closed group of nodes. Thus it is not suitable for decentralized open networks. 
\item In order to confirm the transaction has been committed, the client has to communicate with a super-majority of the nodes to ensure that the transaction is confirmed, making secure transaction verification unscalable.
\end{itemize}

\subsection{Collective Signing}
Within legal infrastructures, we might have several witnesses $W$, and we ask a number of the witnesses $W’$ to verify that something is correct. If one of the witnesses cannot verity the information, we would highlight a problem. Let us say we have a controller on a network, and a number of selected trusted witnesses. Each of the witnesses can then check all of the messages sent by the nodes on the network, and if one of them determines a problem, they can tell the rest of the network. In this respect, every message ($M$) is collectively signed by $W$ witnesses. 

The collective signing (CoSi) algorithm is defined by Syta et al \cite{syta2016keeping}. With CoSi (collective signing), there are four phases involving $P$ participants and where the leader has an index value of zero. Each participant has a private key ($a_i$) and a public key ($A_i=a_iG$, and where $G$ is a base point on an elliptic curve). We then determine the aggregated public key with \cite{syta2016keeping}:

\begin{equation}
A = \sum_{i \in P} A_i
\end{equation}

\textbf{Announcement:} Initially the leader broadcasts a message ($M$) that it wants the participants to sign.

\textbf{Commitment:} Each node $i$ will pick a random scalar ($v_i$) and determines their commitment ($V_i=[v_i]G$). Each commitment is then sent to the leader, who will wait for a specific amount of commitments ($P’$) to be received. The leader then creates a participant bitmask and aggregates all the received commitments:

\begin{equation}
V = \sum_{j \in P'} V_j
\end{equation}

and creates a participation bitmask $Z$. The leader then broadcasts $V$ and $Z$ to the other participants.

\textbf{Challenge:} Each of the participants computes the collective challenge (using the hash function $H$):

\begin{equation}
c = H(V || A || M)
\end{equation}

and send the following back to the leader:

\begin{equation}
r_i = v_i + c \times a_i
\end{equation}

\textbf{Response:} The leader will wait until the participants in $P’$ have sent their responses. Once received, the leader computes the aggregated response:

\begin{equation}
r = \sum_{j \in P'} r_j
\end{equation}

and publishes the signature of the message (M) as:
\begin{equation}
(V,r,Z)
\end{equation}

Each node can then check their own signature value and agree with the leader. 
\bigbreak
\begin{figure*}
  \includegraphics[width=\textwidth]{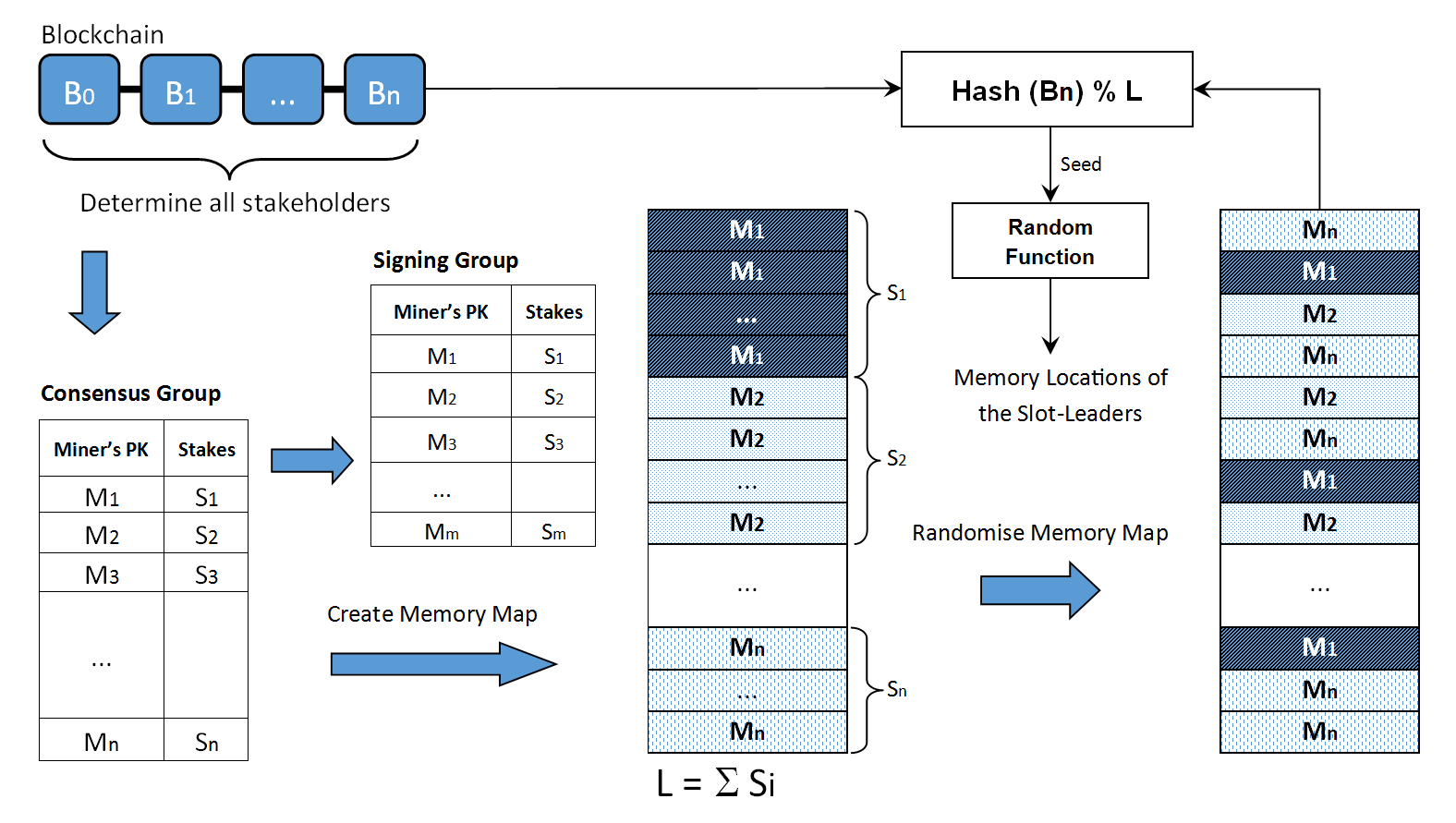}
  \caption{The SklCoin Approach}
  \label{fig:fig1}
\end{figure*}

\section{System Architecture}
SklCoin is designed for untrustworthy networks that can arbitrarily delay, re-order, drop or duplicate messages. We assume the network has a weak synchrony property. The SklCoin system is comprised of a set of $N$ block miners. At any time $t$ a subset of miners $M (t)$ could be faulty or controlled by a malicious attacker. Unlike honest miners, Byzantine miners can attack the system, diverting from our consensus protocol which we call SklCoin.

We assume that we have a group of $n = 3 f + 1$ PBFT nodes. As in PBFT, at any given time, one of these nodes is the leader, who proposes transactions and drives the consensus process. These nodes collectively maintain the blockchain, collecting transactions from clients and adding them in blocks. SklCoin, our consensus protocol, guarantees that only one blockchain history ever exists and that it can never be rolled back or rewritten. The safety and liveness is guaranteed as long as at most $f$ nodes are faulty.   

Subsequent sections describe how to build the SklCoin using the following steps:
\begin{enumerate}
\item We create a mechanism to allow miners to determine the consensus and signing groups in every mining round.
\item We leverage the idea of proof-of-stake to create a mechanism to elect the leader.
\item We leverage the collective signing technique used in ByzCoin to reduce per-round communication complexity to $O(log\ n)$ and reduce typical signature verification complexity from $O(n)$ to $O(1)$.
\end{enumerate}

\subsection{The SklCoin Protocol}
Conventional BFT schemes rely on a well-defined consensus group to guarantee safety and liveness. Sybil attacks can break any open-membership protocol involving security thresholds. For example, PBFT assumes that at most $f$ out of $3f + 1$ members are not honest. To remedy this situation, the proof-of-work mechanism used in Bitcoin, allows only miners who have dedicated resources to become a member of the consensus group. In proof-of-stake, only miners who have shares in the system can become part of that group. Here, we adapt the proof-of-stake mechanism to maintain the “balance of power” within the BFT consensus group over a given fixed-time slot.

The proof-of-stake (PoS) algorithm is the most important part of the SklCoin protocol. It defines how the nodes reach an agreement about the state of the ledger. The idea of PoS is not new and it has been already used in several approaches such as \cite{pos1}\cite{pos2}\cite{pos3}\cite{pos4}. The core idea of PoS is that instead of wasting energy as the case in proof-of-work, a node is elected to generate a new block with a probability proportional to the amount of stake this node has. If a node is elected to generate a new block it will be called "slot leader". 

Similar to the approach in \cite{seijas2018marlowe}, every SklCoin is associated with two fields: a balance and a stake. Balance is the real amount of "coins" that each user has. Similar to any cryptocurrency, any user can send any amount of SklCoins (within this balance), to other users, as well as receive any amount of SklCoins from other users. Stake, on the other hand, gives a user the power to contribute in the evolution of the ledger. For example, only nodes with positive stakes (we call them stakeholders or miners) can participate in running the SklCoin protocol. We differentiate here between two types of transactions: normal transactions where the transferred coins are associated with a balance, and stake transactions where the transferred coins are associated with a stake.

The SklCoin protocol divides the physical time into slots. A slot is a relatively short period of time (e.g. 10sec). Each slot corresponds to one and only one slot-leader which has a sole right to produce one and only one block during his slot. If slot-leader missed their slot (for example, went offline), the right to produce a block is lost until they are elected again. In this case, as will be seen later, SklCoin allows miners to elect the next slot-leader(s).

Coming to a consensus between the miners in the SklCoin protocol boils down to (Fig.\ref{fig:fig1}): 
\begin{enumerate}
\item Determine the consensus group.
\item Select one of the miners to mine a new block in a way that is fair (not biased toward any miners with specific characteristics).
\item Determine the signing group (a subset of the consensus group).
\item Efficiently broadcast the new block to the signing group who verify the content of the block and the eligibility of the slot-leader before adding the new block to their local blockchain. 
\end{enumerate}

\subsection{How is Consensus Group Determined}
We consider here that any miner that has at least one SklCoin with stake is eligible to be a slot-leader. The idea here is that, at the beginning of each block-mining round (time-slot), each miner can determine who is the slot-leaders in the coming slots. Ideally, at the beginning of the time-slot, each miner has to know only who will be the slot-leader of that slot. However, as some miners may become unavailable, a timeout mechanism is used that allows the miners to know the slot-leaders of the coming slots in case the slot-leader of the current slot went offline. 

Similar to the "fair lottery" idea, the selection mechanism should guarantee that any stakeholder can become a slot-leader. However, it should also guarantee that the more stakes the miner has, the more its chance to be elected. In that respect, each miner parses all stake transactions available in all blocks starting from the genesis block and calculates the stakes of each stakeholder (miner). This way, the stakeholders will be sorted according to their appearance in the blockchain. This list of stakeholders can be cached in the miner's machine and it can be updated when a new block is appended to the blockchain. The result is that all miners will end up with the same ordered list of stakeholders (consensus group) along with their number of stakes.

\subsection{How is Slot-leader Selected}
Each miner follows the following steps:

\textbf{Step 1:} After the new block has been added to their local blockchain, each miner calculates the hash of the last added block; we call that hash the "common seed" (CS). This CS is used as a seed to a random number generator to generate a sequence of random numbers with a length equals to the total amount of coins staked into the network. For instance, if the amount of total coins staked to the network is 100, then the random number generator must create a sequence of 100 numbers.

\textbf{Step 2:} Each miner maintains a memory map where each memory location corresponds to a specific stakeholder's (miner) public key. Suppose we have $N$ stakeholders. Each stakeholder $i$ has $S_i$ stakes. Therefore, the memory map will have $L = \sum S_i$ locations in the memory. Each location contains the stakeholder's public key. The more stakes the stakeholder has the more memory locations they will have, and the more chance they will be elected. The question now is how to distribute these locations in memory. We need to ensure that all miners end up with creating the same memory map in their local machines. Each miner will use the same CS, produced in Step 2, as an input to a deterministic pseudo-random numbers function which returns random numbers between 1 and $L$. Because all miners use the same CS, all miners will end up with the same sequence of random numbers. This sequence of numbers specifies the memory locations of each of the $L$ stakes. The first random number of that sequence specifies the memory location where we add the public key of the stakeholder that corresponds to the first stake available in the list of Step 1. The second random number specifies the memory location where we add the public key of the stakeholder that corresponds to the second stake available in the list, and so on. If the random number specifies a memory location that has been already taken, we move to the next random number. The result is that all miners will generate the same memory map that contains a randomly distributed list of stakeholders' public keys.

\textbf{Step 3:} At this point we need to have a mechanism that allows miners to know who is the winning leader in the current time slot, and the next time slots in case the current leader went offline. Each miner calculates the value ($CS\ modulo\ L$) to be used as a seed to a random number generator which produces a series of random numbers between $1$ and $L$, and which determine the sequence of next leaders. The first randomly generated number specifies the memory location of the winning leader in the current time-slot. If for some reason the winning miner went offline, other miners will wait till the time of the current slot times out, and then the next randomly generated number is used to determine the next leader, and so on.
\subsection{How is Signing Group Determined}
In order to ensure the scalability and feasibility of SklCoin, a maximum number of miners $M$ should be selected to form the signing group. That group is responsible of validating and agreeing on the block proposed by the slot-leader. For this objective, the list of miners (resulted in Step 1) is sorted according to their stakes and the top $M$ is selected to become the signing group of the current slot. 

The slot-leader prepares a new block and broadcasts it to the signing group as will be seen in the next section. If the slot leader produces a block that gets included in the chain, they receive a block reward equal to the total fees of all transactions. 

\subsection{Agreeing on the New Block}
The slot-leader starts generating a new block which will contain the transactions it received. In the standard PBFT protocol, there exist two phases: 1) the pre-prepare phase in which the leader obtains attestations from a super-majority quorum of signing group members (two-thirds) that the leader’s proposal is safe and consistent with all previously committed history, and 2) the commit phase in which the leader obtains attestations from a super-majority that all the signing members witnessed the successful result of the prepare phase and they committed to remember the decision.

We adopt the collective signing approach taken in \cite{kogias2016enhancing}. To implement the pre-prepare phase, the slot-leader announces the first collective signature (CoSi) round. The resulted CoSi from the prepare phase provides a proof-of-acceptance of a proposed block of transactions by the slot-leader. In the second commit round, the slot-leader announces the proof-of-acceptance to all members, who then validate it and collectively sign the block’s hash to produce a collective commit signature on the block. This way a Byzantine leader cannot rewrite history or double-spend.

Validating the slot-leader's proposal comprises two points: 1) The content of the block in the proposal must be correct. That includes the list of transactions and that the block is cryptographically linked with the last block. 2) The originator of the proposal is the correct slot-leader of the current time slot. After validating the new block, each member of the signing group (including the slot-leader) adds that block to its local blockchain. In addition, the slot-leader broadcasts the new block along with the collective signature to the miners of the consensus group who did not participate in the signing group. The result is that all miners end up with the same blockchain state.

\section{Discussion}
It is inherently difficult to analyze such complex system, and therefore we only consider some common attacks and leave remaining more formal security analysis for future work.

\subsection{Attacks}
It is important to ensure that our consensus model functions correctly in normal as well as adversarial conditions. In a permission-less environment, the number of miners is expected to be large, and these miners are anonymous and untrusted since any node is allowed to join the network. Consensus mechanisms for such environments have to be resilient the following attacks:

\textbf{Sybil attacks:} Sybil attacks on a blockchain network can allow a single or group of entities to generate several (millions) online identities that they control to influence and manipulate the consensus process. Having such dominance allows these entities to confirm the transactions and blocks as per their rules or to include double-spend transactions. 

In SklCoin, generating multiple (millions) identities by the attacker is useless because: 1) participating in the mining process (consensus group) requires the attacker to stake some coins into the network; it is unlikely that miners play against their interest, 2) the probability of selecting the leader is a function of the amount of stakes rather than the number of identities, and 3) since participating in the signing group depends on the number of stakes, generating multiple identities does not help the attacker to be elected as a member of the signing group.

\textbf{Selfish and deliberately malicious nodes:} selfish mining occurs when the normally honest miners are incentivized to support the attacker and join in carrying out an attack. In this attack, the attacker performs erratic mining, at the cost of his short term revenue by maintaining a separate private blockchain in parallel to the Bitcoin blockchain. He selectively publishes many blocks all at once, forcing rest of the network to discard their blocks and ultimately losing revenue. 

It is true that the attacker can be part of the consensus group by having as low as one SklCoin (stake); however, because the chance of being elected as a leader is proportional to the numbers of stakes, the attacker will have a slim chance of mining new blocks. But what if the attacker has been elected to be the leader? In that case, SklCoin reduces the risk by 1) electing the attacker for only one time slot, and 2) any attempt to add a malicious block will be noticed by the witnesses.

\textbf{Nothing-at-stake attack}: This attack refers to the case when a miner (aka forger) forges on every possible fork. In fact, proof-of-stake is susceptible to forks being created accidentally or maliciously giving the miner the chance to mine on more than one fork simultaneously, thus, enabling double spending attacks. This is impossible with PoW as a miner will consume power mining when mining on two forks simultaneously. However, with PoS, forging on multiple chains costs the miner nothing and will be rewarded  no matter which fork wins. In SklCoin, there is no chance for the network to generate multiple forks, thus, eliminating the risk of mounting the nothing-at-stake attack.

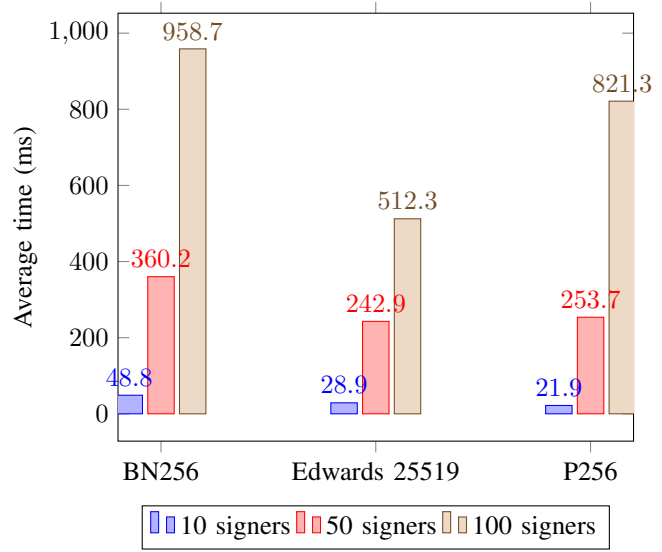
\begin{figure}
\begin{tikzpicture}  
\begin{axis}[
    ybar,
    enlargelimits=0.1,
    legend style={at={(0.5,-0.15)}, anchor=north,legend columns=-1},
    ylabel={Average time (ms)},
    symbolic x coords={BN256,Edwards 25519,P256},
    xtick=data,
    nodes near coords,
    nodes near coords align={vertical},
    ]
\addplot coordinates {(BN256,48.8) (Edwards 25519, 28.9) (P256, 21.9)};
\addplot coordinates {(BN256,360.2) (Edwards 25519, 242.9) (P256,253.7)};
\addplot coordinates {(BN256,958.7) (Edwards 25519, 512.3) (P256, 821.3)};
\legend{10 signers, 50 signers, 100 signers}
\end{axis}

\end{tikzpicture}
\caption{The performance evaluation - Average time}
\label{fig2}
\end{figure}

An experiment on an AMD Phenom II running at 2.8~GHz produced the results defined in Fig.\ref{fig2}. This uses  differing elliptic curves of: BN256 \cite{barreto2005pairing}; Edwards 25519 (with BlakeSHA256) \cite{hisil2008twisted}; and P256 (with BlakeSHA256) \cite{adalier2015efficient}, and uses the Golang Kyber library \cite{kyber}. The experiment uses a number of signers for each run. We create a new key pair for each signer, create a new mask for the signing process, and then create an aggregated commitment and mask. Finally the commitments and responses are created for each signer. It can be seen that Edwards 25519 and P256 produces faster signing than BN256. The efficiency of scale can be seen with the Edwards 25519 curve. There are some worries around the current security levels of BN256 \cite{kim2016extended}, where the security level may drop to 96 bits of security.

\subsection{Limitations of the SklCoin protocol}
Recently, it turned out that, according to \cite{OnTheSecurity}, CoSi was not proved secure. The authors introduced mBCJ, a secure two-round multi-signature scheme. Their results show that mBCJ is only marginally less efficient than CoSi, so that any protocol based on the insecure CoSi scheme should instead be built on the provably secure mBCJ scheme. In the SklCoin approach, replacing the collective signature scheme does not affect the approach itself and the plan is to replace CoSi with the mBCJ scheme as the next step.

In addition, SklCoin does not have a mechanism that prevents a leader miner from producing a block that will not get included in the blockchain. That mechanism is necessary to stop miners from producing blocks that won’t get included in the main chain. This could happen when one or more malicious entities are trying to slow the throughput rate by producing fake or malicious blocks, or not producing blocks at all. However, SklCoin reduces this risk by: 1) requiring the leaders to have stakes, and 2) the leader is not allowed to mine more than one block during the leader-slot. 

\subsection{Advantages of the SklCoin protocol}
The advantages include:

\begin{itemize}
\item The leader is selected from the consensus group which is formed from all miners who have stakes. In order to ensure scalability, the leader is responsible of choosing the signing group which is a subset of the consensus group. 
\item No Blockchain Fork: A blockchain fork can result in different nodes in the system converging on different blocks as being part of the blockchain. Unlike Bitcoin where temporary forks may exist due to network latencies, SklCoin has immediate finality characteristics i.e. once the transaction is included in the block, it is confirmed and will not be rolled back.
\item Transaction rate is higher with platforms that can confirm transactions immediately and reach consensus fast. Wait time between blocks can be significant e.g., up to 10 minutes using Bitcoin’s difficulty tuning scheme. Whereas, in SklCoin one block is produced in each time slot. The time slot is a configurable parameter and it could depends on the network size. Unlike PoW approaches which are probabilistic and have to spend significant amount of time solving a cryptographic puzzle, SklCoin supports higher transaction rate because of a faster mechanism for leader election. 
\item Scalability of the protocol: The scalability of SklCoin comes from: 1) the mechanism of electing the leader is fast and does not depend on the number of available miners, and 2) electing a subset of the consensus group to become the signing group. 
\end{itemize}

\section{Conclusion}
SklCoin is an attempt to solve the scalability and low-throughput rate in open decentralized blockchain systems. By leveraging the ideas of proof-of-stake and the collective signature, SklCoin provides strong consistency and prevents common attacks on the consensus and mining system such as blockchain forks and selfish-mining. SklCoin allows open membership in the consensus group i.e any node can become a member of the miners as long it gets some stakes. The consensus group and signing group are determined by the miners in each time slot. Future work includes creating a penalty mechanism to punish the malicious leaders who is either not adding any blocks or adding malicious blocks, and finding fairer mechanism of choosing the signing group that does not depend only on the number of stakes. In addition, more formal security analysis of the proposed protocol considering how it is able to mitigate some known attacks in the permissionless setting is needed. 

\bibliographystyle{IEEEtran}
\bibliography{references}

\begin{thebibliography}{10}
\providecommand{\url}[1]{#1}
\csname url@samestyle\endcsname
\providecommand{\newblock}{\relax}
\providecommand{\bibinfo}[2]{#2}
\providecommand{\BIBentrySTDinterwordspacing}{\spaceskip=0pt\relax}
\providecommand{\BIBentryALTinterwordstretchfactor}{4}
\providecommand{\BIBentryALTinterwordspacing}{\spaceskip=\fontdimen2\font plus
\BIBentryALTinterwordstretchfactor\fontdimen3\font minus
  \fontdimen4\font\relax}
\providecommand{\BIBforeignlanguage}[2]{{%
\expandafter\ifx\csname l@#1\endcsname\relax
\typeout{** WARNING: IEEEtran.bst: No hyphenation pattern has been}%
\typeout{** loaded for the language `#1'. Using the pattern for}%
\typeout{** the default language instead.}%
\else
\language=\csname l@#1\endcsname
\fi
#2}}
\providecommand{\BIBdecl}{\relax}
\BIBdecl

\bibitem{Un-blockchain}
T.~Dinh, R.~Liu, M.~Zhang, G.~Chen, B.~Ooi, and J.~Wang, ``Untangling
  blockchain: A data processing view of blockchain systems,'' \emph{IEEE
  Transactions on Knowledge and Data Engineering}, vol.~PP, 08 2017.

\bibitem{technical-survey}
F.~Tschorsch and B.~Scheuermann, ``“bitcoin and beyond: A technical survey on
  decentralized digital currencies,'' \emph{IEEE Communications Surveys
  Tutorials}, vol.~18, no.~3, pp. 2084--2123, 2016.

\bibitem{kogias2016enhancing}
E.~K. Kogias, P.~Jovanovic, N.~Gailly, I.~Khoffi, L.~Gasser, and B.~Ford,
  ``Enhancing bitcoin security and performance with strong consistency via
  collective signing,'' in \emph{25th $\{$USENIX$\}$ Security Symposium
  ($\{$USENIX$\}$ Security 16)}, 2016, pp. 279--296.

\bibitem{nxtcoin}
S.~Popov, ``A probabilistic analysis of the nxt forging algorithm,''
  \emph{LEDGER}, vol.~1, 2016.

\bibitem{Castro}
\BIBentryALTinterwordspacing
M.~Castro and B.~Liskov, ``Practical byzantine fault tolerance,'' in
  \emph{Proceedings of the Third Symposium on Operating Systems Design and
  Implementation}, ser. OSDI '99.\hskip 1em plus 0.5em minus 0.4em\relax
  Berkeley, CA, USA: USENIX Association, 1999, pp. 173--186. [Online].
  Available: \url{http://dl.acm.org/citation.cfm?id=296806.296824}
\BIBentrySTDinterwordspacing

\bibitem{Understanding}
A.~Baliga, ``Understanding blockchain consensus models,'' 2017.

\bibitem{liu2018scalable}
J.~Liu, W.~Li, G.~O. Karame, and N.~Asokan, ``Scalable {B}yzantine consensus
  via hardware-assisted secret sharing,'' \emph{IEEE Transactions on
  Computers}, vol.~68, no.~1, pp. 139--151, 2018.

\bibitem{miller2016honey}
A.~Miller, Y.~Xia, K.~Croman, E.~Shi, and D.~Song, ``The honey badger of bft
  protocols,'' in \emph{Proceedings of the 2016 ACM SIGSAC Conference on
  Computer and Communications Security}.\hskip 1em plus 0.5em minus 0.4em\relax
  ACM, pp. 31--42.

\bibitem{crain2018dbft}
T.~Crain, V.~Gramoli, M.~Larrea, and M.~Raynal, ``Dbft: Efficient leaderless
  {B}yzantine consensus and its application to blockchains,'' in \emph{2018
  IEEE 17th International Symposium on Network Computing and Applications
  (NCA)}.\hskip 1em plus 0.5em minus 0.4em\relax IEEE, 2018, pp. 1--8.

\bibitem{Bitcoin-NG}
\BIBentryALTinterwordspacing
I.~Eyal, A.~E. Gencer, E.~G. Sirer, and R.~V. Renesse, ``Bitcoin-ng: A scalable
  blockchain protocol,'' in \emph{13th {USENIX} Symposium on Networked Systems
  Design and Implementation ({NSDI} 16)}.\hskip 1em plus 0.5em minus
  0.4em\relax Santa Clara, CA: {USENIX} Association, Mar. 2016, pp. 45--59.
  [Online]. Available:
  \url{https://www.usenix.org/conference/nsdi16/technical-sessions/presentation/eyal}
\BIBentrySTDinterwordspacing

\bibitem{ohta1999multi}
K.~Ohta and T.~Okamoto, ``Multi-signature schemes secure against active insider
  attacks,'' \emph{IEICE Transactions on Fundamentals of Electronics,
  Communications and Computer Sciences}, vol.~82, no.~1, pp. 21--31, 1999.

\bibitem{Fiat-Shamir}
\BIBentryALTinterwordspacing
``Fiat-shamir heuristic,'' 2019. [Online]. Available:
  \url{https://en.wikipedia.org/wiki/Fiat-Shamir\_heuristic}
\BIBentrySTDinterwordspacing

\bibitem{Solida}
\BIBentryALTinterwordspacing
I.~Abraham, D.~Malkhi, K.~Nayak, L.~Ren, and A.~Spiegelman, ``{Solida: A
  Blockchain Protocol Based on Reconfigurable Byzantine Consensus},'' in
  \emph{21st International Conference on Principles of Distributed Systems
  (OPODIS 2017)}, ser. Leibniz International Proceedings in Informatics
  (LIPIcs), J.~Aspnes, A.~Bessani, P.~Felber, and J.~Leit{\~a}o, Eds.,
  vol.~95.\hskip 1em plus 0.5em minus 0.4em\relax Dagstuhl, Germany: Schloss
  Dagstuhl--Leibniz-Zentrum fuer Informatik, 2018, pp. 25:1--25:19. [Online].
  Available: \url{http://drops.dagstuhl.de/opus/volltexte/2018/8640}
\BIBentrySTDinterwordspacing

\bibitem{LearnBitcoin}
K.~Kulkarni, in \emph{Learn Bitcoin and Blockchain: Understanding Blockchain
  and Bitcoin Architecture to Build Decentralized Applications}.\hskip 1em plus
  0.5em minus 0.4em\relax Packt, 2018.

\bibitem{syta2016keeping}
E.~Syta, I.~Tamas, D.~Visher, D.~I. Wolinsky, P.~Jovanovic, L.~Gasser,
  N.~Gailly, I.~Khoffi, and B.~Ford, ``Keeping authorities" honest or bust"
  with decentralized witness cosigning,'' in \emph{2016 IEEE Symposium on
  Security and Privacy (SP)}.\hskip 1em plus 0.5em minus 0.4em\relax Ieee,
  2016, pp. 526--545.

\bibitem{pos1}
A.~Kiayias, A.~Russell, B.~David, and R.~Oliynykov, ``Ouroboros: A provably
  secure proof-of-stake blockchain protocol,'' in \emph{Advances in Cryptology
  -- CRYPTO 2017}, J.~Katz and H.~Shacham, Eds.\hskip 1em plus 0.5em minus
  0.4em\relax Cham: Springer International Publishing, 2017, pp. 357--388.

\bibitem{pos2}
W.~Li, S.~Andreina, J.-M. Bohli, and G.~Karame, ``Securing proof-of-stake
  blockchain protocols,'' in \emph{Data Privacy Management, Cryptocurrencies
  and Blockchain Technology}, J.~Garcia-Alfaro, G.~Navarro-Arribas,
  H.~Hartenstein, and J.~Herrera-Joancomart{\'i}, Eds.\hskip 1em plus 0.5em
  minus 0.4em\relax Cham: Springer International Publishing, 2017, pp.
  297--315.

\bibitem{pos3}
\BIBentryALTinterwordspacing
F.~Saleh, ``Blockchain without waste: Proof-of-stake,'' 2019. [Online].
  Available: \url{http://dx.doi.org/10.2139/ssrn.3183935}
\BIBentrySTDinterwordspacing

\bibitem{pos4}
\BIBentryALTinterwordspacing
T.~Duong, L.~Fan, and H.-S. Zhou, ``2-hop blockchain: Combining proof-of-work
  and proof-of-stake securely,'' 2017. [Online]. Available: \url{. https:
  //eprint.iacr.org/2016/716.}
\BIBentrySTDinterwordspacing

\bibitem{seijas2018marlowe}
P.~L. Seijas and S.~Thompson, ``Marlowe: Financial contracts on blockchain,''
  in \emph{International Symposium on Leveraging Applications of Formal
  Methods}.\hskip 1em plus 0.5em minus 0.4em\relax Springer, 2018, pp.
  356--375.

\bibitem{barreto2005pairing}
P.~S. Barreto and M.~Naehrig, ``Pairing-friendly elliptic curves of prime
  order,'' in \emph{International Workshop on Selected Areas in
  Cryptography}.\hskip 1em plus 0.5em minus 0.4em\relax Springer, 2005, pp.
  319--331.

\bibitem{hisil2008twisted}
H.~Hisil, K.~K.-H. Wong, G.~Carter, and E.~Dawson, ``Twisted edwards curves
  revisited,'' in \emph{International Conference on the Theory and Application
  of Cryptology and Information Security}.\hskip 1em plus 0.5em minus
  0.4em\relax Springer, 2008, pp. 326--343.

\bibitem{adalier2015efficient}
M.~Adalier \emph{et~al.}, ``Efficient and secure elliptic curve cryptography
  implementation of curve p-256,'' in \emph{Workshop on Elliptic Curve
  Cryptography Standards}, vol.~66, 2015.

\bibitem{kyber}
DEDIS, ``{DEDIS Advanced Crypto Library for Go},''
  \url{https://github.com/dedis/kyber}, 2019, [Online; accessed 20-Oct-2019].

\bibitem{kim2016extended}
T.~Kim and R.~Barbulescu, ``Extended tower number field sieve: A new complexity
  for the medium prime case,'' in \emph{Annual International Cryptology
  Conference}.\hskip 1em plus 0.5em minus 0.4em\relax Springer, 2016, pp.
  543--571.

\bibitem{OnTheSecurity}
M.~Drijvers, K.~Edalatnejad, B.~Ford, E.~Kiltz, J.~Loss, G.~Neven, and
  I.~Stepanovs, ``On the security of two-round multi-signatures,'' 05 2019, pp.
  1084--1101.

\end{thebibliography}

\end{document}